\documentclass[aps,prl,twocolumn,superscriptaddress,preprintnumbers,showpacs,floatfix]{revtex4}
\usepackage{graphics,colordvi}
\def\eq#1\en{\begin{equation} #1 \end{equation}}

\def\eqa#1\ena{\begin{eqnarray} #1 \end{eqnarray}}

\usepackage{graphicx,color}

\begin{document}

\title{Constraining noncommutative field theories with holography}

\author{Raul Horvat}
\author{Josip Trampeti\'{c}}
\affiliation{Physics Division, Rudjer Bo\v skovi\' c Institute, Zagreb, Croatia}

\date{\today}

\begin{abstract}
An important window to quantum gravity phenomena in low energy
noncommutative (NC) quantum field theories (QFTs) gets represented by a
specific form of UV/IR mixing. Yet another important window to quantum
gravity, a holography, manifests itself in effective QFTs as a distinct
UV/IR connection. In matching these two principles, a useful relationship
connecting the UV cutoff $\Lambda_{\rm UV}$, the IR cutoff  
$\Lambda_{\rm IR}$ and the scale
of noncommutativity $\Lambda_{\rm NC}$, can be obtained. We show that an
effective QFT endowed with both principles may not be capable  to fit
disparate experimental bounds simultaneously, like the muon $g-2$ and the 
masslessness of the photon. Also, the constraints from the muon $g-2$ 
preclude any possibility to observe the
birefringence of the vacuum coming from objects at
cosmological distances. On the other hand, in
NC theories without the UV completion, where the perturbative aspect of the
theory (obtained by truncating a power series in 
$ \Lambda_{\rm NC}^{-2}$)
becomes important, a heuristic estimate of the region where the
perturbative expansion is well-defined $E/ \Lambda_{\rm NC} \lesssim 1$, gets
affected when holography is applied by providing the energy of the system
$E$ a $\Lambda_{\rm NC}$-dependent 
lower limit. This may affect models which try to infer the scale
$\Lambda_{\rm NC}$ by using data from low-energy experiments.   
\end{abstract}

\newpage

\maketitle

Quantum field theories (QFTs) constructed on noncommutative (NC) spacetime:
\begin{equation}
[x_{\mu}, x_{\nu}] = i\theta_{\mu \nu} \;,
\label{theta}
\end{equation}
have received a great deal of interest lately mainly because of the possible
appearance of such spacetimes in string theory 
\cite{Connes:1997cr,Seiberg:1999vs,Douglas:2001ba,Szabo:2001kg,Chu:2005ev}. In these QFTs 
noncommutativity is
characterized by a real antisymmetric matrix $\theta_{\mu \nu}$ 
of dimension of length squared,
which in the context of string theory reflects the properties of the
background. At tree level a QFT formulated with (\ref{theta}) should switch to its
commutative relative whenever the momenta of the field quanta are lowered 
below $\theta^{-1/2}$. In contrast, at loop level the inherent
nonlocality of the full theory shows up in the UV/IR mixing phenomenon \cite{Minwalla:1999px},
meaning that switching to ordinary theories may occur at much lower momenta,
depending on the ultimate UV cutoff in the theory. Figuratively speaking, the
effect describes the linear growth of the size of a particle with its
momentum, showing thus unambiguously its quantum-gravity origin \cite{Matusis:2000jf}.   

The phenomenon of UV/IR  
mixing \cite{Minwalla:1999px} is best understood by examining the behavior
of the (nonplanar) loop graphs with the ordinary product of fields replaced 
by the Moyal star($\star$)-product (see e.g.,\cite{Douglas:2001ba,Szabo:2001kg}). This results in 
phase factors depending on the
virtual momenta of internal loops \cite{Filk:1996dm}. In a theory without UV completion
($\Lambda_{\rm UV} \rightarrow \infty$) these phase factors, although efficient in
damping out the high-energy part of the graphs, becomes together
inefficient to control the vanishing momenta, i.e., the original UV
divergences reappear as IR divergences. On the other hand, 
in presence of a finite $\Lambda_{\rm UV}$ no one sort of divergence 
will appear since the said phase factors effectively transform the 
highest energy
scale ($\Lambda_{\rm UV}$) into the lowest one ($\Lambda_{\rm IR}$). The 
theory thus
becomes an effective  QFT with the UV and the IR cutoffs obeying a
relationship,
\begin{equation}
\Lambda_{\rm UV} \Lambda_{\rm IR} \sim \Lambda_{\rm NC}^2 \;,
\label{Lambda}
\end{equation}     
where the scale of noncommutativity is heuristically introduced as
$\Lambda_{\rm NC}^{-2} \sim |\theta|$. To a good approximation the theory 
boils down to an ordinary (local) commutative theory 
where the crucial information from quantum gravity
is brought in by the relationship (\ref{Lambda}).
 
The power of the UV/IR connection (\ref{Lambda}) is best seen 
heuristically by inspecting the
photon self-energy, where explicit results in the NC approach in presence of
a finite $\Lambda_{\rm UV}$ do exist both in the case of the external momentum 
above and below $\Lambda_{\rm IR}$ \cite{AlvarezGaume:2003mb,Jaeckel:2005wt,Abel:2006wj}. 
We know that in a commutative setting the integral is badly UV divergent
if we regulate in the most naive way, giving
\cite{Peskin-Schroder}
\footnote{Although a truly NC part of
the photon self-energy has a different tensorial part, $\tilde{k}^{\mu}\tilde{k}^
{\nu}/\tilde{k}^2$, with $\tilde{k}^{\mu} = \theta^{\mu \nu} k_{\nu}$,
heuristically it has the same momentum dependence as $g^{\mu \nu}$, so it can
be omitted from this qualitative comparison. See for details 
\cite{Hayakawa:1999yt}. Also, note that in gauge non-invariant regularization
schemes, UV/IR mixing in NC field theories is best understood
\cite{AlvarezGaume:2003mb}.}  
\begin{equation}
\Pi(k)^{\mu \nu} \propto g^{\mu \nu} \Pi(k) \sim 
g^{\mu \nu} \Lambda_{\rm UV}^2 \,.
\label{Pi}
\end{equation}
From (\ref{Lambda}) one expect the same result to be valid also 
in the explicit NC approach for $k \stackrel{>}{\sim} \Lambda_{\rm IR}$, as
confirmed in \cite{Abel:2006wj}. For $k \lesssim \Lambda_{\rm IR}$
(\ref{Lambda}) is of course ineffective but in the deep IR regime one uses to invoke a
standard quadratic decoupling of the massive species to get 
\begin{equation}
\Pi(k) \sim (k^2/\Lambda_{\rm IR}^2) \Lambda_{\rm UV}^2 \;.
\label{Pik}
\end{equation} 
For a generic non-SUSY theory, this agrees, by applying (\ref{Lambda}), with
$\tilde{k}^2 \Lambda_{\rm UV}^4$ from \cite{Abel:2006wj}. When SUSY is softly broken 
$\Lambda_{\rm UV}^4 \rightarrow \Delta M_{\rm SUSY}^2 \Lambda_{\rm UV}^2$ 
\cite{AlvarezGaume:2003mb,Jaeckel:2005wt,Abel:2006wj},
where $\Delta M_{\rm SUSY}^2$ is the supertrace of the mass matrix. 

In the present paper we shall impose on 
the above QFT an extra requirement
from the realm of quantum gravity: the holographic principle. Then we
would like to test such a `beefed up' theory against experimental data. We
show that the latter requirement shows up as an additional and distinct
UV/IR correspondence. We find that additional restrictions imposed by
holography become insurmountable barriers when
fitting disparate experimental data.

For an effective QFT in a box of size $L$ (providing an IR
cutoff $\sim \Lambda_{\rm IR}^{-1}$) the entropy scales extensively,
$S_{\rm QFT} \sim L^3 \Lambda_{\rm UV}^3$, and therefore there is always 
a sufficiently large volume
for which $S_{\rm QFT}$ would exceed the absolute Bekenstein-Hawking bound
$S_{\rm BH}\sim L^2 M_{Pl}^2$, where $M_{Pl}$ is the Planck mass. 
Thus, considerations for the maximum possible entropy
suggest that ordinary QFT may not be valid for arbitrarily large volumes,
unless the UV and the IR cutoffs obey a constraint, 
$L\Lambda_{\rm UV}^3 \lesssim M_{Pl}^2$. However, 
at saturation, this bound means that an
effective QFT should also be capable to describe systems containing
black holes, since it necessarily includes many states with Schwarzschild
radius $R_S$ much larger than the box size. There are however arguments for why an
effective QFT appears unlikely to
provide an adequate description of any system containing black holes
\cite{'tHooft:1993gx,Lowe:1995ac}. So, ordinary QFT may not be
valid for much smaller volumes, but would apply provided a more stringent
constraint is obeyed \cite{Cohen:1998zx}
\begin{equation}
\Lambda_{\rm UV}^3 \Lambda_{\rm IR}^{-3} \lesssim M_{Pl}^{3/2} \Lambda_{\rm
IR}^{-3/2} \sim S_{\rm BH}^{3/4} \;.
\label{hologr}
\end{equation}

For a NC theory described by (\ref{Lambda}) and obeying the holographic requirement
(\ref{hologr}), additional constraints can be derived
\begin{eqnarray}
\Lambda_{\rm IR}
&\stackrel{>}\sim&
\Lambda_{\rm NC} \Big(\frac{\Lambda_{\rm NC}}{M_{Pl}} \Big)^{1/3} 
\label{IRNC}\\
\Lambda_{\rm UV}
&\lesssim&
\Lambda_{\rm NC} \Big(\frac{M_{Pl}}{\Lambda_{\rm NC}} \Big)^{1/3}
\label{UVNC}
\end{eqnarray}
which represent, in addition to (\ref{Lambda}), a further connection between the
parameters describing the field theory ($\Lambda_{\rm UV}$ 
and $\Lambda_{\rm IR}$) 
and that describing spacetime ($\Lambda_{\rm NC})$  

One may however justifiably object that our bounds (5-7), as they stand, do
not reflect possible modifications due to NC black hole thermodynamics.
Following \cite{Banerjee:2008gc,{Nicolini:2005vd}}, such effects indeed do
arise due to the fuzziness of space induced by the space-space component of
the uncertainty relation (1). Consequently, various thermodynamic entities,
like the mass of the black hole
and the Schwarzschild radius get modified in a spherically symmetric,
stationary NC Schwarzschild spacetime
\cite{Banerjee:2008gc,{Nicolini:2005vd}}. The Bekenstein-Hawking entropy
$S_{BH}$ also  receives corrections when $\theta \neq 0$. For fixed $\theta$ and
an arbitrary-sized black hole, the Schwarzschild radius in NC settings cannot
be obtained in a closed form, but useful relations can be obtained in the
large radius regime $R_{S}^{2}/4 \theta >> 1$. It was shown
\cite{Banerjee:2008gc} that neglecting the  usual logarithmic correction
at $\theta =0$ (otherwise unimportant in our
case), and at the leading order in the parameter  $R_{S}^{2}/4 \theta$, both
the NC horizon area $A$ and the NC black hole entropy follow the same 
functional change
as a function of $\theta$, such that the exact functional form of the
(usual) commutative area law, $S_{BH}^{NC} = AM_{Pl}^2/4$, stays preserved.
This means that NC thermodynamical laws are a NC deformation 
of the usual laws. Thus our bounds (5-7) are unaffected in NC settings. One
can furthermore check whether the above field-theoretical setup fits within
the given regime  $R_{S}^{2}/4 \theta >> 1$. One can easily show that the
constraint $R_{S}^{2}/4 \theta >> 1$, with the aid of Eq. (2), boils down to
$\Lambda_{\rm UV}/\Lambda_{\rm IR} >> (L/R_{S})^2$. The ${\it rhs}$ of the
latter constraint is always $\stackrel{>}\sim 1$, by noting (as discussed before) 
that any effective field theory approach should not describe states already
collapsed to a black hole. Thus, we are left with $\Lambda_{\rm
UV}/\Lambda_{\rm IR} >> 1$, which is nothing but the consistency constraint
for any ordinary QFT.

Equipped with these relationships we seek for further constraints on
$\Lambda_{\rm UV}$, $\Lambda_{\rm IR}$ and $\Lambda_{\rm NC}$ by testing the
theory against experimental data. First we seek for further 
constraints on $\Lambda_{\rm UV}$, $\Lambda_{\rm
IR}$ and $\Lambda_{\rm NC}$ by considering the muon $g-2$. Of essence here
is to notice that a contribution from radiative corrections which otherwise
would tend to zero when both $\Lambda_{\rm UV} \rightarrow \infty$,
$\Lambda_{\rm IR} \rightarrow 0$, now yields a finite answer because
$\Lambda_{\rm UV}$ and $\Lambda_{\rm IR}$ should obey a nontrivial constraint
from the UV/IR mixing (\ref{Lambda}) as well as the holographic constraints 
(\ref{hologr}). We have,
\begin{equation}
\Delta(g_\mu-2) \sim \frac{\alpha}{\pi} \left [\left
(\frac{m_{\mu}}{\Lambda_{\rm UV}} \right )^2 + \left (\frac{\Lambda_{\rm
 IR}}{m_{\mu}} \right )^2 \right ] \;.
\label{DeltaIRUV}
\end{equation}
Since we are no more able to pick out $\Lambda_{\rm UV}$ independently from
$\Lambda_{\rm IR}$ the contribution (\ref{DeltaIRUV}) becomes nonzero. Obviously, this  
correction of the quantum-gravitational origin 
greatly surpasses the usual Planck-scale correction. 

By applying (\ref{Lambda}) and
assuming first $\Lambda_{\rm NC} \stackrel{>}{\sim}  m_{\mu}$ 
one arrives at 
\begin{equation}
\Delta(g_\mu-2)_{\rm IR} \sim \frac{\alpha}{\pi} \left (\frac{\Lambda_{\rm IR}}{m_{\mu}}
\right )^2 \;.
\label{DeltaIR} 
\end{equation} 
From the report of the muon E821 anomalous magnetic moment measurements at
BNL \cite{Bennett:2006fi} we know that 
\begin{equation}
\frac{g_{\mu} - 2}{2}(\rm{Exp - SM}) = (22 -26) \times 10^{-10} \;.
\label{Expg-2}
\end{equation}
In turn, this and the holographic constraint (\ref{IRNC}) on $\Lambda_{\rm IR}$
implies 
\begin{equation}
m_{\mu} \lesssim \Lambda_{\rm NC} \lesssim 0.1 \;\rm  TeV \;.
\label{01TeV}
\end{equation}
Also one obtains $\Lambda_{\rm IR} \lesssim 10^{-1} \;~\rm MeV$ and 
$10^5 \;~\rm MeV \lesssim \Lambda_{\rm
UV} \lesssim 10^5 \;~\rm TeV$. 

In the opposite regime, $\Lambda_{\rm NC} \lesssim 
m_{\mu}$, one gets
\begin{equation}
\Delta(g_\mu-2)_{\rm UV} \sim \frac{\alpha}{\pi} \Big(\frac{m_{\mu}}{\Lambda_{\rm UV}}
\Big)^2 \;,
\label{DeltaIR} 
\end{equation}
from where the constraint (\ref{UVNC}) and Eq. (\ref{Expg-2}) give
\begin{equation}
10^{-4} \rm \;~ MeV \lesssim \Lambda_{\rm NC} \lesssim m_{\mu} \;,
\label{10-4}
\end{equation}
together with $\Lambda_{\rm UV} \stackrel{>}{\sim}  10^2 \;~\rm  GeV$ 
and $ 10^{-1} \;~\rm  MeV \stackrel{>}{\sim} \Lambda_{\rm IR} 
\stackrel{>}{\sim} 10^{-13} \;~\rm \ MeV$. Note that our
constraints (\ref{01TeV}) and (\ref{10-4}) are the exceptional ones, 
providing for the first time
the upper bound for the scale of noncommutativity,
\begin{equation}
\Lambda_{\rm NC} \lesssim  {\cal O}(0.1)\;~ \rm TeV. 
\label{0.1TeV}
\end{equation}

The other NC approaches (without UV completion and holography, whether based on
the truncated theory or not) when confronting experiments may yield only
a lower limit on $\Lambda_{\rm NC}$. The best limits yielded 
$\Lambda_{\rm NC} \stackrel{>}{\sim} {\cal O}(100) \;~\rm  TeV$ \cite{Horvat:2010sr}. This
striking mismatch could signal how strongly may the outcome be influenced by
the NC UV/IR mixing and holography. Also,
there is an illuminating argument \cite{AmelinoCamelia:2002au} that 
when calculating loop effects any NC approach based on
the truncated-$\theta$ expansion can reliably approximate the full
theory only in the region below $\Lambda_{\rm IR}$, where the result
crucially depends on the UV completion of our model, i.e., the region 
where our
effective theory is deprived of predictive power.

Note that lowering 
the UV scale down to the dark energy scale of
the universe at present, $\Lambda_{\rm UV} \sim 10^{-3}\;~ \rm eV$, is not
allowed by our constraints (\ref{01TeV}) and (\ref{10-4}). 
The present theoretical setup 
is hence not capable to shed light on the dark energy/cosmological constant
problem \cite{Nobbenhuis:2006yf}.

In the following we show how strong theoretical and experimental
constraints as outlined above would actually make our NC setup incapable to
account for other experimental data. We shall concentrate on the UV/IR
mixing in the polarization tensor for a pure U(1) NC gauge theory, 
\cite{Jaeckel:2005wt,Abel:2006wj}. 
When SUSY is softly broken, a
nontrivial dispersion relation induces a sizable Lorentz symmetry
violating mass term for the photon, if the photon momentum $k \stackrel{>}{\sim} 
\Lambda_{\rm IR}$ \cite{Abel:2006wj}. Since masslessness of the photon is well tested
up to at least $1 \;~ \rm GeV$, one requires
\begin{equation}
\Lambda_{\rm IR}  \stackrel{>}{\sim} 1 \;~ \rm GeV \;,
\end{equation}        
which is strikingly inconsistent with both cases (\ref{01TeV}) and (\ref{10-4}). In a different
regime, $\Lambda_{\rm IR} \stackrel{>}{\sim} \rm 1 \;~ TeV$, the same dispersion
relation would induce the birefringence effect for gamma rays 
having energies $\sim 1 \;~ \rm TeV$, and reaching us from
astrophysical sources at the cosmological distances. Such an effect 
would alter the light speed to $\approx c(1 - \Delta n)$. Again, the
constraints from the muon $g-2$ preclude any possibility for the NC-induced
birefringence for high energy gamma rays. Still, one may be curious to see
the impact of our basic constraints (\ref{Lambda}), 
(\ref{IRNC}) and (\ref{UVNC}) on the said birefringence
effect. Employing the strongest limit on $\Delta n$ to date \cite{Kostelecky:2006ta}, attaining the
level of $10^{-37}$ and coming from the recent measurements of linear
polarization in gamma rays from the two gamma-ray bursts at 
$z \stackrel{>}{\sim} 0.1$, one obtains \cite{Abel:2006wj}
\begin{equation}
(10^{18} \rm GeV)^2 \,\Lambda_{\rm UV}^2 \,\Lambda_{\rm NC}^{-4}\;~ \lesssim \;~10^{-3}\;.
\end{equation}   
Thus, based solely on the birefringence constraint, one sees,  from Eqs. (\ref{Lambda}),
(\ref{IRNC}) and (\ref{UVNC}), that 
the Planck-scale noncommutativity, 
$\Lambda_{\rm UV} \sim \Lambda_{\rm NC} \stackrel{>}{\sim} M_{Pl}$, is preferred.  

Although (as already noted above) the truncated $\theta$-expansion is not
likely to be a reliable approximation of the full theory when considering
loop effects, i.e. considering 
the expansion in $k\theta \ell$ with $k(\ell)$ playing the
role of external (loop) momentum, there have been, on the other hand, 
a great deal of attempts
undertaken to constrain $\Lambda_{\rm NC}$ by using scattering tree-level processes 
within such an approach \cite{Behr:2002wx}. 
In that case the truncated $\theta$-expanded theory can
give a good approximation of the full theory if heuristically 
$E/\Lambda_{\rm NC} \lesssim 1$, 
where $E$ is the energy of the system. The less $E/\Lambda_{\rm NC}$, the
better the reliability of the approximation. The holographic 
restrictions (\ref{IRNC}) and (\ref{UVNC})  
can however give a consequential constraint to this statement by providing $E$ a
lower limit. Indeed, $E \stackrel{>}{\sim} \Lambda_{\rm IR}$ requires
\begin{equation}
E \;~\stackrel{>}{\sim} \;~ \Lambda_{\rm NC} \left (\frac{\Lambda_{\rm NC}}{M_{Pl}}
\right)^{1/3} \;,
\end{equation}
by (\ref{IRNC}). For instance, if the inferred $\Lambda_{\rm NC}$ (obtained 
by comparing a given scattering
process to the data) is around $\rm 10 \;~ TeV$, then 
$E \stackrel{>}{\sim} \rm 100 \;~ MeV$.
This obviously can affect placing limits on $\Lambda_{\rm NC}$ when 
using high-precision low-energy experiments, notably with atomic constraints 
on $\Lambda_{\rm NC}$ \cite{Chaichian:2000si}.

In conclusion, we have shown that information from two distinct realms
of quantum gravity may not be readily
manifested when imposed on an effective QFT. In a sense, the UV/IR mixing
from noncommutativity and holography do not go hand in hand when implemented
in the field-theoretical approach. Although an ultimate UV cutoff is
expected to be capable of providing a good approximation to the effect of UV
completion, owing to the string theory realization of both
noncommutativity/holography, still a proven lack of universality in NC 
theories \cite{Minwalla:1999px,Khoze:2000sy} 
may cause a failure of the theory to fit simultaneously 
disparate experimental constraints. Or simply an information from two
distinct UV/IR mixings, (\ref{Lambda}) and (\ref{hologr}) respectively, 
when implemented on models at low energy scales, becomes 
more overabundant and scrambled than with the either UV/IR mixing taken
separately.  
    
\section*{Acknowledgment}
We would like to thank L. Alvarez-Gaume,
J. Ellis and P. Minkowski for discussions.
The work of R.H. and J.T. are supported by 
the Croatian  Ministry of Science, Education and Sports 
under Contract Nos. 0098-0982930-2872 and 0098-0982930-2900, respectively. 
%J.T. would like to thank P. Schupp for useful discussions,
%and W. Hollik at MPI, Muenchen for the hospitality.
The work of J.~T. is in part supported by the EU (HEPTOOLS) 
project under contract MRTN-CT-2006-035505.

%{\bf Acknowledgment. } This work was supported by the Ministry of Science,
%Education and Sport
%of the Republic of Croatia under contract No. 098-0982887-2872.

\end{document}